\documentclass[aps,preprint,preprintnumbers,amsmath,amssymb,nofootinbib]{revtex4}
\usepackage{amssymb}
\usepackage{lscape}
\usepackage{amsmath}
\usepackage{rotating}
\usepackage{bm}
\usepackage{graphicx}
\graphicspath{ {images/} }
\begin{document}
\title{Geometric back-reaction in pre-inflation from relativistic quantum geometry}
\author{$^{2}$ Marcos R. A. Arcod\'{\i}a\footnote{E-mail address: marcodia@mdp.edu.ar},  $^{1,2}$ Mauricio Bellini
\footnote{E-mail address: mbellini@mdp.edu.ar} }
\address{$^1$ Departamento de F\'isica, Facultad de Ciencias Exactas y
Naturales, Universidad Nacional de Mar del Plata, Funes 3350, C.P.
7600, Mar del Plata, Argentina.\\
$^2$ Instituto de Investigaciones F\'{\i}sicas de Mar del Plata (IFIMAR), \\
Consejo Nacional de Investigaciones Cient\'ificas y T\'ecnicas
(CONICET), Mar del Plata, Argentina.}

\begin{abstract}
The pre-inflationary evolution of the universe describes the beginning of the expansion from a static initial state, such that the Hubble parameter is initially zero, but increases to an asymptotic constant value, in which it could achieve a de Sitter (inflationary) expansion. The expansion is driven by a background phantom field. The back-reaction effects at this moment should describe vacuum geometrical excitations, which are studied with detail in this work using Relativistic Quantum Geometry.
\end{abstract}
\maketitle

\section{Introduction}

The inflationary model is a very tested description of how the universe can provide a physical mechanism to
generate primordial energy density fluctuations on cosmological scales \cite{infl}, below Planckian scales. During this stage, the primordial
scalar perturbations drove the seeds of large scale structure which had then gradually formed today's galaxies. This is being
tested in current observations of cosmic microwave background (CMB)\cite{smoot}. These fluctuations are today larger than a thousand times the size of a typical galaxy, but during inflation they
were very much larger than the size of the causal horizon\cite{prd96}. According to this scenario, the almost constant potential depending of a minimally coupled to
gravity inflation field $\phi$, called the inflaton, caused the accelerated expansion of the very early universe. During this epoch, the potential energy density was dominant, so that the kinetic energy  can be neglected. This is known as the slow-roll condition for the inflaton field dynamics.
In this framework the problem of nonlinear (scalar) perturbative corrections to the metric has been studied in \cite{muk}.

Geometrodynamics\cite{Wh} is a picture of general relativity that study the evolution of the space-time geometry. The significant advantages of geometrodynamics, usually come at the expense of manifest local Lorentz symmetry\cite{R}. During the 70's and 80's decades a method of quantization was developed in order to deal with some unresolved problems of quantum field theory in curved spacetimes\cite{pru}. In this context, recently we have introduced a new method to study the scalar perturbations of the metric in a non-perturbative manner\cite{Mio} by introducing Relativistic Quantum Geometry (RQG). This formalism is non-perturbative and serves to describe the dynamics of the geometric departure of a background Riemannian spacetime with the help of a quantum geometrical scalar field\cite{rb,rb1,RB}. The dynamics of the geometrical scalar field is defined on a Weyl-integrable manifold that preserves the gauge-invariance under the transformations of the Einstein's equations, that involves the cosmological constant. Our approach is different to quantum gravity. The natural way to construct quantum gravity models is to apply quantum field theory methods to the theories of classical gravitational fields interacting with matter. In spite of numerous efforts the general problems of quantum gravity still remain unsolved. Our approach is different because our subject of study is the
dynamics of the geometrical quantum fields. This dynamics is obtained from the Einstein-Hilbert action, and not by using the standard effective action used in various models of quantum gravity\cite{od}. There are no non-linearities or high-derivative problems in the dynamical description, so our formalism is much easier to apply to different physical systems like inflation\cite{Mio}, or pre-inflation. This primordial epoch is of relevant interest in cosmology and deserves a detailed study. Presently, we cannot understand completely the first epoch of the universal evolution. How the universe begun to expand and how was the first stage of this evolution?
The theory that describes this epoch is called pre-inflation\cite{Kur}. The existence of a pre-inflationary epoch with fast-roll of the inflaton field would introduce an infrared depression in the primordial power spectrum. This depression might have left an imprint in the CMB anisotropy\cite{GR}. It is supposed that during pre-inflation the universe begun to expand from some Planckian-size initial volume to thereafter pass to an inflationary epoch. In this framework RQG should be very useful when we try to study the evolution of the geometrical back-reaction effects given that we are dealing with Planckian energetic scales, and back-reaction effects should be very intense at these scales.

In this work we shall describe a model in which the universe begins to expand from nothing (the initial energy density is null). The Hubble parameter is initially zero and thereafter increases to an asymptotic constant value, in which it could achieve a de Sitter (inflationary) expansion. In our model the background expansion is driven by a phantom scalar field $\phi$, in which the equation of state of the universe during pre-inflation is $P_{pi}/\rho_{pi} < -1$\cite{phantom}. The back-reaction effects at this moment should describe vacuum geometrical excitations, which are the main subject of study in this work.

\section{Relativistic Quantum Geometry: the structure of spacetime in an expanding universe}

We shall consider a metric tensor in the Riemannian manifold with null covariant derivative
(we denote with a semicolon the Riemannian-covariant derivative): $\Delta \bar{g}_{\alpha\beta}=\bar{g}_{\alpha\beta;\gamma} \,dx^{\gamma}=0$,
such that the Weylian\cite{Weyl} covariant derivative $ \bar{g}_{\alpha\beta|\gamma} = \theta_{\gamma}\,\bar{g}_{\alpha\beta}$, described with respect to the Weylian
connections \footnote{To simplify the notation we shall denote $\theta_{\alpha} \equiv \theta_{,\alpha}$, where the comma denotes the partial derivative. Furthermore, we shall exalt with a " bar " the quantities represented on the Riemannian background manifold. }
\begin{equation}\label{gama}
\Gamma^{\alpha}_{\beta\gamma} = \left\{ \begin{array}{cc}  \alpha \, \\ \beta \, \gamma  \end{array} \right\}+ \theta^{\alpha} \bar{g}_{\beta\gamma} ,
\end{equation}
is nonzero
\begin{equation}\label{gab}
\delta \bar{g}_{\alpha\beta} = \bar{g}_{\alpha\beta|\gamma} \,dx^{\gamma} = -\left[\theta_{\beta} \bar{g}_{\alpha\gamma} +\theta_{\alpha} \bar{g}_{\beta\gamma}
\right]\,dx^{\gamma}.
\end{equation}
In the case of an expanding universe, the Riemannian manifold will be described by the background geometry characterized with a FRW metric. Of course, all the variations with respect to the expanding background are in the Weylian geometrical representation. As was demonstrated in \cite{RB} the Einstein tensor can be written as
\begin{equation}\label{ein}
\bar{G}_{\alpha\beta} = {G}_{\alpha\beta} + \theta_{\alpha ; \beta} + \theta_{\alpha} \theta_{\beta} + \frac{1}{2} \,g_{\alpha\beta}
\left[ \left(\theta^{\mu}\right)_{;\mu} + \theta_{\mu} \theta^{\mu} \right] \equiv G_{\alpha\beta} - \bar{g}_{\alpha\beta} \Lambda ,
\end{equation}
and we can obtain the semi-Riemannian invariant (the cosmological constant) $\Lambda$
\begin{equation}\label{p}
\Lambda = -\frac{3}{4} \left[ \theta_{\alpha} \theta^{\alpha} + \bar{\Box} \theta\right].
\end{equation}
Notice that $\Lambda(\theta,\theta^{\alpha})$ is a Riemannian invariant, but not a Weylian invariant.
Hence, one can define a geometrical Weylian quantum action
${\cal W} = \int d^4 x \, \sqrt{-\bar{g}} \, \Lambda(\theta,\theta^{\alpha})$, such that the dynamics of the geometrical field, after imposing $\delta
W=0$, is described by the Euler-Lagrange equations which take the form
\begin{equation}\label{q}
\bar{\nabla}_{\alpha} \Pi^{\alpha} =0, \qquad {\rm or} \qquad \bar\Box\theta=0.
\end{equation}
The canonical momentum components are $\Pi^{\alpha}\equiv -{3\over 4} \theta^{\alpha}$ and the relativistic quantum algebra is given by\cite{RB}
\begin{equation}\label{con}
\left[\theta(x),\theta^{\alpha}(y) \right] =- i \Theta^{\alpha}\, \delta^{(4)} (x-y), \qquad \left[\theta(x),\theta_{\alpha}(y) \right] =
i \Theta_{\alpha}\, \delta^{(4)} (x-y),
\end{equation}
with $\Theta^{\alpha} = i \hbar\, \bar{U}^{\alpha}$ and $\Theta^2 = \Theta_{\alpha}
\Theta^{\alpha} = \hbar^2 \bar{U}_{\alpha}\, \bar{U}^{\alpha}$ for the Riemannian components of velocities $\bar{U}^{\alpha}$.

\section{Pre-inflation and back-reaction}

One of the most important paradigms in the cosmology consists in providing an explanation of how was the initial moment of the expansion of the universe. This implies providing a model of how the universe begun its expansion before the inflationary accelerated expansion with a Hubble parameter very close to a constant and $P_i/\rho_i \gtrsim -1$. A possible scenario is pre-inflation, in which the
Hubble parameter is initially null, to thereafter increase to a asymptotic constant value. During the beginning of the expansion the universe has a equation of state with $P_{pi}/\rho_{pi} < -1$, which implies that the expansion is driven by a minimally coupled to gravity scalar phantom field $\phi$. During pre-inflation the action is
\begin{equation}
{\cal I} = \int_{V} d^4 x \sqrt{|\bar{g}|} \,  \left[\frac{\bar{R}}{2\kappa} + \frac{\lambda}{2} \dot\phi^2 - V(\phi)\right].
\end{equation}
where $\kappa = 8 \pi G$, $G$ is the gravitational constant, $\sqrt{|\bar{g}|} = a^3(t)$ is the volume of the manifold $\mathcal{M}$, and $\bar{g}_{\mu\nu} = {\bf diag}[1, -a^2, -a^2, -a^2]$ are the components of the diagonal tensor metric. With the aim to describe pre-inflation, we shall use $\lambda=-1$, which describes the dynamics of a fast-rolling phantom field. However, this epoch would be followed by an inflationary expansion driven by the slow-rolling inflaton field, for which the dynamics is obtained when $\lambda=1$.
Here, $\phi(t)$ is the background solution that describe the dynamics of a isotropic and homogeneous background metric that characterizes a semi-Riemannian manifold.
\begin{equation}\label{phi}
\ddot\phi + 3 H \dot\phi  + \lambda V'(\phi)=0,
\end{equation}
where $V(\phi)$ is the potential and the prime denotes the derivative with respect to $\phi$. The semi-Riemannian (background) Einstein equations are
\begin{equation}\label{ein}
\bar{G}_{\alpha\beta} \equiv \bar{R}_{\alpha\beta} - \frac{1}{2} \,\bar{g}_{\alpha\beta}\,\bar{R} = -\kappa\, \bar{T}_{\alpha\beta},
\end{equation}
where the components of the background stress tensor are $\bar{T}_{\alpha\beta} = \frac{\delta {\cal \bar{L}}}{\delta \bar{g}^{\alpha\beta}} - \bar{g}_{\alpha\beta} {\cal \bar{L}} $. For a background FRW metric the Einstein equations result
\begin{eqnarray}
3 H^2 &=& \kappa \rho_{pi}= \kappa \left[ \lambda \frac{\dot\phi^2}{2} + V(\phi)\right],  \label{e1}\\
-\left(3 H^2 + 2 \dot{H}\right) &=& \kappa\, P_{pi} = \kappa \left[ \lambda \frac{\dot\phi^2}{2} - V(\phi)\right]. \label{e2}
\end{eqnarray}
From the two Einstein equations we obtain that $\dot{\phi} = - \frac{ 2 H'}{\kappa \lambda}$, and the time dependent potential can be written as a function of the Hubble parameter and its
time derivative:
\begin{equation}
V(t) = \frac{1}{\kappa} \left[3H^2 + \dot{H}\right].
\end{equation}
This expression can be re-written taking into account the $\phi$-dependence
\begin{equation}
V(\phi) = \frac{1}{\kappa} \left[3 H^2(\phi) - \frac{2}{\kappa\lambda} \left(H'\right)^2 \right].
\end{equation}

\subsection{The pre-inflationary model with a phantom field}

We consider a model in which the Hubble parameter which is initially zero, and tends asymptotically to $\left.H\right|_{t \gg 1/(\sqrt{2}A)} \rightarrow H_0$.
\begin{equation}\label{h}
H(t) = H_0 \, \tanh{\left[2\,H_0\, t\right]},
\end{equation}
where the cosmological constant is related to $H_0$: $\Lambda=3 H^2_0$. The scale factor of the universe during this stage is
\begin{equation}\label{aa}
a(t)= \frac{a_0}{\left[1-\tanh^2{\left(2\, H_0\, t\right)}\right]^{1/4}},
\end{equation}
with $a_0 =H^{-1}_0$. Notice that this solution describes an universe in which $\dot H >0$. In other words, the model describes an universe which begin to
expanding since an initial scale factor $a(t=0) \equiv H^{-1}_0$. Furthermore, the Hubble parameter increases super-exponentially from a null value to an asymptotically constant value.
The scalar potential can be written as a function of $t$
\begin{equation}\label{v1}
V(t)= \frac{H^2_0}{\kappa} \left[ \tanh^2{(2 H_0 t)} + 2\right],
\end{equation}
so that for sufficiently large times, we obtain that $\left. V(t)\right|_{t\rightarrow \infty} \rightarrow \frac{3 H^2_0}{\kappa}$. From the Einstein equations (\ref{e1}) and
(\ref{e2}), we obtain the time dependence of $\dot{\phi}$
\begin{equation}
\dot{\phi} = \frac{2H_0}{\sqrt{\kappa}} \left[ 1- \tanh^2{(2 H_0 t)}\right]^{1/2}.
\end{equation}
Using the fact that $V'=\frac{\dot{V}}{\dot{\phi}}$ in the equation of motion (\ref{phi}), we obtain the time dependence of the background scalar field
\begin{equation}
\phi(t)= \frac{2}{\sqrt{\kappa}} \, \arctan{\left(e^{2 H_0 t}\right)} - \frac{\pi}{2\sqrt{\kappa}},
\end{equation}
where $0 \leq \phi \leq \frac{\pi}{2\sqrt{\kappa}}$. Notice that the phantom field increases during pre-inflation. Therefore, if we use this expression in the equations (\ref{h}) and (\ref{v1}), we obtain the $\phi$-dependence of the Hubble parameter and the scalar potential
\begin{eqnarray}
H(\phi) &=& H_0 \left[
1- 2 \cos^2\left[\frac{\sqrt{\kappa}}{2} \left(\phi+\frac{\pi}{2\sqrt{\kappa}}\right)\right]\right], \\
V(\phi) &=& \frac{H^2_0}{\kappa} \left\{\left[
1- 2 \cos^2\left[\frac{\sqrt{\kappa}}{2} \left(\phi+\frac{\pi}{2\sqrt{\kappa}}\right)\right]\right]^2 +2\right\},
\end{eqnarray}
such that $V(\phi(t=0)]=\frac{2 H^2_0}{\kappa} \leq V(\phi) \leq V[\phi(t\rightarrow\infty)] = \frac{3H^2_0}{\kappa}$. Notice that
$\rho_{pi}(t=0) = 0$, so that in this model the universe is created from nothing.

\subsection{Back-reaction effects in pre-inflation}

The geometrical scalar field $\theta$ can be expressed as a Fourier expansion
\begin{equation}\label{four}
\theta(\vec{x},t) = \frac{1}{(2\pi)^{3/2}} \int \, d^3k \, \left[ A_k \, e^{i \vec{k}.\vec{x}} \xi_k(t) + A^{\dagger}_k \, e^{-i \vec{k}.\vec{x}} \xi^*_k(t) \right],
\end{equation}
where $A^{\dagger}_k$ and $A_k$ are the creation and annihilation operators. From the point of view of the metric tensor, an example in power-law inflation can be illustrated by
\begin{equation}\label{met1}
g_{\mu\nu} = {\rm diag}\left[ e^{2\theta}, - a^2(t) e^{-2\theta}, - a^2(t) e^{-2\theta}, - a^2(t) e^{-2\theta}\right],
\end{equation}
where the scale background scale factor $a(t)$ is given by (\ref{aa}). The quantum volume of the manifold described by (\ref{met1}) is $V_q= a^3(t) e^{-2\theta}= \sqrt{-\bar{g}} \,e^{-2\theta}$. The dynamics for $\theta$ is governed by the equation
\begin{equation}
\ddot\theta + 3 \frac{\dot{a}}{a} \dot\theta - \frac{1}{a^2} \nabla^2 \theta =0,
\end{equation}
and the momentum components are $\Pi^{\alpha}\equiv -{3\over 4} \theta^{\alpha}$, so that the relativistic quantum algebra is given by the expressions (\ref{con}) with
co-moving relativistic velocities $U^0=1$, $U^i=0$.

Furthermore, as was calculated in a previous work\cite{Mio}, the variation of the energy density fluctuations is given by the expression
\begin{equation}\label{de}
\frac{1}{\bar{\rho}} \frac{\delta \bar{\rho}}{\delta S} = - 2 \theta_0 =  -2 \dot{\theta},
\end{equation}
such that $\dot{\theta} \equiv \left<B|\dot\theta^2|B\right>^{1/2}$. To understand what is the line element $S$ in a quantum context, we can define the operator
\begin{equation}
\check{x}^{\alpha}(t,\vec{x}) = \frac{1}{(2\pi)^{3/2}} \int d^3 k \, \check{e}^{\alpha} \left[ b_k \, \check{x}_k(t,\vec{x}) + b^{\dagger}_k \, \check{x}^*_k(t,\vec{x})\right],
\end{equation}
where $b^{\dagger}_k$ and $b_k$ are the creation and annihilation operators of spacetime, such that $\left< B \left| \left[b_k,b^{\dagger}_{k'}\right]\right| B  \right> = \delta^{(3)}(\vec{k}-\vec{k'})$ and $\check{e}^{\alpha}=\epsilon^{\alpha}_{\,\,\,\,\beta\gamma\delta} \check{e}^{\beta} \check{e}^{\gamma}\check{e}^{\delta}$,
where $\epsilon^{\alpha}_{\,\,\,\,\beta\gamma\delta}$ are the Levi-Civita symbols. In this framework, we must understand that the exact differential related to a coordinate $x^{\alpha}$
\begin{equation}
dx^{\alpha} \left. | B \right> =  \bar{U}^{\alpha} dS \left. | B \right>= \delta\check{x}^{\alpha} (x^{\beta}) \left. | B \right> ,
\end{equation}
is the eigenvalue that results when we apply the operator $ \delta\check{x}^{\alpha} (x^{\beta}) $ on the background quantum state $ \left. | B \right> $, defined as a Fock space on the Riemannian manifold. The Weylian line element is given by
\begin{equation}
dS^2 \, \delta_{BB'}=\left( \bar{U}_{\alpha} \bar{U}^{\alpha}\right) dS^2\, \delta_{BB'} = \left< B \left|  \delta\check{x}_{\alpha} \delta\check{x}^{\alpha}\right| B'  \right>.
\end{equation}
Hence, the differential Weylian line element $dS$ provides the displacement of the quantum trajectories with respect to the "classical" (Riemannian) ones: $d\bar{S}^2 =
\bar{g}_{\alpha\beta} dx^{\alpha} dx^{\beta}$.

\subsection{Quantization of modes}

\begin{figure}[h]
\includegraphics[width=1.0\textwidth]{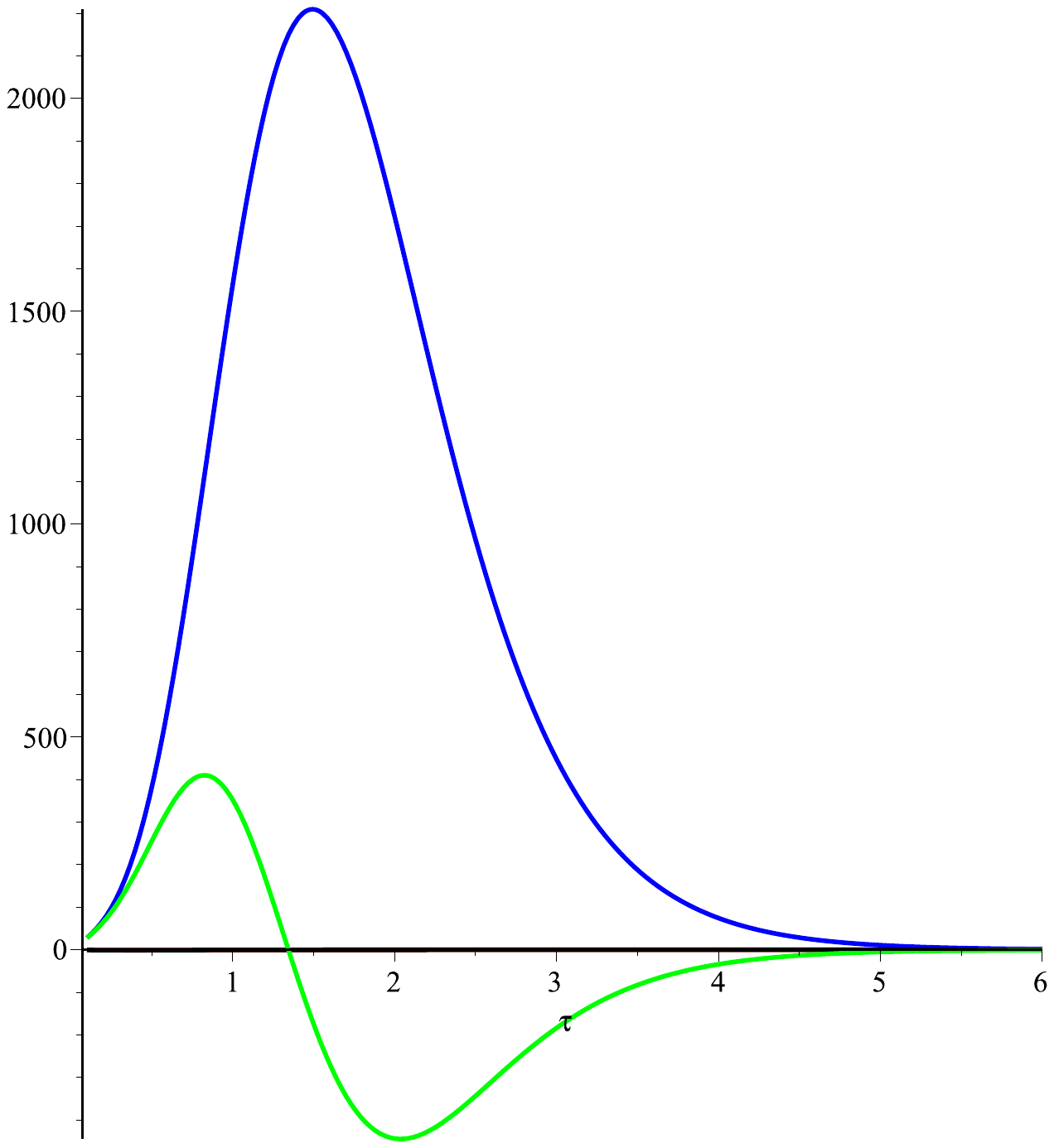}\caption{Contributions to $\left<B|\dot\theta^2|B\right>_{k_n}$ drawn for $a_0=G^{1/2}$, due to the modes $k_1=1.414213562\,i$ (red),  $k_3=1.394736996\,i$ (blue), $k_5=4.034677761$ (black), and $k_7=-4.082914929+0.6506152090\,i$ (green). }\label{f1}
\end{figure}

The equation of motion for the modes $\xi_k(t)$ is
\begin{equation}\label{xi}
\ddot\xi_c(t) + 3 \frac{\dot a}{a} \dot\xi_c(t) + \frac{k^2}{a(t)^2} \xi_c(t)=0.
\end{equation}
The annihilation and creation operators $B_{k}$ and
$B_{k}^{\dagger}$ satisfy the usual commutation algebra
\begin{equation}\label{m5}
\left[A_{k},A_{k'}^{\dagger}\right]=\delta ^{(3)}(\vec{k}-\vec{k'}),\quad
\left[A_{k},A_{k'}\right]=\left[A_{k}^{\dagger},A_{k'}^{\dagger}\right]=0.
\end{equation}
Using the commutation relation (\ref{m5}) and the Fourier
expansions (\ref{four}), we obtain the normalization condition for
the modes. For convenience we shall re-define the dimensionless time: $\tau =b\,t$, where $b=\sqrt{\frac{2\Lambda}{3}}=\frac{1}{a_0}$, so that the normalization condition for  $\xi_{c}(\tau)$ is
\begin{equation}\label{m6}
\xi_{k}(t) \frac{d{\xi}^*_{k}(\tau)}{d\tau} - \xi^*_{k}(\tau) \frac{d{\xi}_{k}(\tau)}{d\tau} = i
\,\left(\frac{a_0}{a(\tau)}\right)^{3},
\end{equation}
where the asterisk denotes the complex conjugated.
The general solution for the modes $\xi_k(\tau)$, is
\begin{eqnarray}
\xi_k(\tau) &=& C_1\, \frac{\sinh{(\tau)}}{\sqrt{2\cosh^2{(\tau)}-1}}  \times {\rm Hn}\left[-1,\frac{k^2-1}{4};0,\frac{1}{2},\frac{3}{2},\frac{1}{2};-\tanh^2{(\tau)}\right] \nonumber \\
&+& C_2\, \frac{\cosh{(\tau)}}{\sqrt{2\cosh^2{(\tau)}-1}} \times  {\rm Hn}\left[-1,\frac{k^2+1}{4};-\frac{1}{2},0,\frac{1}{2},\frac{1}{2};-\tanh^2{(\tau)}\right] , \nonumber \\
\label{solu}
\end{eqnarray}
where ${\rm Hn}[a,q;\alpha,\beta,\gamma,\delta;z]=\sum^{\infty}_{j=0} c_j\, z^j$ is the Heun function. Since the Heun functions are written as infinite series, we can make a series expansion in both sides of (\ref{m6}), in order to obtain the restrictions for the coefficients $C_1$ and $C_2$, and the wavenumber values $k$. The polynomial expansion of $\xi_{k}(t) \left[{\xi}^*_{k}(\tau)\right]' - \xi^*_{k}(t) \left[{\xi}_{k}(\tau)\right]' = i
\,\left(\frac{a_0}{a(\tau)}\right)^{3}$, can be written as a series expansion
\begin{equation}
\xi_{k}(t) \left({\xi}^*_{k}(\tau)\right)' - \xi^*_{k}(t) \left({\xi}_{k}(\tau)\right)' - i\,\left(\frac{a_0}{a(\tau)}\right)^{3} = \sum^{\infty}_{N=1} f_N(k)\, \tau^N=0,
\end{equation}
where $f_N(k^{(N)}_n)=0$, for each $N$. To simplify the notation we are denoting the $\tau$-derivative with a {\it prime}. There are $2N$ modes for each $N$-th order of the expansion, which cames from the
roots of each equation. These roots provide us with the discrete quantum modes becoming from the quantization of $\theta$.
From the zeroth order of the expansion (in $\tau$), we obtain
that $C_2=-i\,C_1/2$. Hence, we shall choose $C_1=1$ results that $C_2=-i/2$ in the general solution (\ref{solu}).
The first eight terms of the series, are
\begin{eqnarray}
\sum^{\infty}_{N=1} f_N(k)\, \tau^N & = &
\frac{2}{3}\,i \left( {k}^{2}+2 \right) {\tau}^{2}+4/3\,i \left( {k}^{2}+2 \right) {\tau}^{3}+  \frac {2}{45}\,i \left( -43\,{k}^{2}-95+3\,{k}^{4} \right) {\tau}^{4}\nonumber \\
& + &{\frac {4}{15}}\,i \left( -21\,{k}^{2}-45+{k}^{4} \right) {\tau}^{5} + \frac {1}{945}\,i \left( 3667
\,{k}^{2}+8996-474\,{k}^{4}+12\,{k}^{6} \right) {\tau}^{6} \nonumber \\
& + & \frac {2}{315}\,i \left( 2263\,{k}^{2}+5248-214\,{k}^{4}+4\,{k}^{6} \right) \tau^{7} \nonumber \\
& + &  \frac {1}{14175}\,i \left( -99563\,{k}^{2}-268995+17513\,{
k}^{4}-820\,{k}^{6}+10\,{k}^{8} \right) {\tau}^{8} + ... = 0.\nonumber \\ \label{sum}
\end{eqnarray}
From each $k$-dependent polynomial we obtain the roots, which provide us the permitted modes that guarantee the quantization of $\theta$. There are
infinite discrete permitted modes.
The expectation value for $\dot\theta^2$ on the quantum state $\left.|B\right>$ calculated on the background semi-Riemannian hypersurface, is
\begin{equation}\label{dot}
\left<B|\dot\theta^2|B\right> = \frac{a^2_0}{(2\pi^2)}\sum^{\infty}_{n=1} \left(k_n\, k^*_n\right)\,\left[ \left[{\xi}_{k_n}(\tau)\right]'\left[{\xi}^*_{k_n}(\tau)\right]' \right],
\end{equation}
such that $\left({\xi}_{k_n}(\tau)\right)'$ is $\tau$-derivative of $\xi_{k_n}$
evaluated at $k=k_n$: $\left.\left({\xi}_{k}(\tau)\right)'\right|_{k=k_n}$ and $k_n$\footnote{$k^*_n$ is the complex conjugate}, are the complex roots of the polynomials $f_N(k)=0$, in (\ref{sum}). The expression (\ref{dot}) can be alternatively written for each mode $k_n$, as
\begin{equation}\label{dot1}
\left<B|\dot\theta^2|B\right>_{k_n} = \frac{a^2_0}{(2\pi^2)}\, \left(k_n\, k^*_n\right)\,\left[ \left[{\xi}_{k_n}(\tau)\right]'\left[{\xi}^*_{k_n}(\tau)\right]' \right],
\end{equation}
which takes into account the contribution of each $k_n$-mode in $\left<B|\dot\theta^2|B\right>$. We see that the first modes have roots in $k_{1,2}= \pm \sqrt{2}\, i$. The modes for these
roots have the same contribution in the expression for $\left<B|\dot\theta^2|B\right>_{k_{1,2}}$. The modes of the second polynomial in (\ref{sum}) are the same. from the modes of the third
polynomial come from the roots of $3\,{k}^{4}-43\,{k}^{2}-95=0$ are $k_{3,4,5,6}=1.394736996\,i, -1.394736996\,i, 4.034677759, -4.034677759$. The modes of the fourth polynomial have roots in $k_{7,8,9,10}=1.399977069\,i, -1.399977069\,i, 4.791652721, -4.791652721$. In the figure (\ref{f1}) we were drawn the contributions of the modes $k_1$ (red), $k_3$ (blue), $k_5$ (black) and $k_7$ (green), to $\left<B|\dot\theta^2|B\right>_{k_n}$, for $a_0=G^{1/2}$. Notice that all the contributions tends asymptotically to zero for few Planckian times ($t_p\simeq 10^{-43}\, sec$). In other words, the excitations of the background (i.e., the Riemannian vacuum), are significative at the moment of the big bang, but decrease to zero when $\dot{H}/H^2 \rightarrow 0$. This corresponds just to the approximation to the de Sitter (inflationary) regime.

\section{Final comments}

We have studied back-reaction effects in a pre-inflationary universe using RQG. This formalism makes possible the nonperturbative treatment of the vacuum fluctuations of the spacetime,
by making a displacement from a semi-Riemannian to a Weylian one. In this framework the Einstein equations are exactly valid on the Riemannian manifold, but the quantum effects are described on the Weylian one by the field $\theta$. In the Weylian manifold the cosmological constant is not an invariant, but a Lagrangian density $\Lambda(\theta,\theta^{\alpha})$ with which we define the quantum action ${\cal W}$. The dynamics of the geometrical field $\theta$ is that of a free scalar field and describes the dynamics of the geometrical quantum fluctuations with respect to the Riemannian (classical) background.
When we apply this formalism to the pre-inflationary scenario, which describes the beginning of the universal expansion from nothing, we obtain that the modes of the geometrical field $\theta$ are discrete, but infinite in number. The contribution of some of these to the variation of energy density, were drawn in the figure (\ref{f1}). Notice that all the modes's contributions become asymptoticaly null for $\tau \gg 1$. Other distinctive characteristic of these modes is that they are not unstable, as in the case of the modes of $\theta$ during inflation\cite{Mio}. A subsequent study on how to describe the transition from pre-inflation to inflation remains pendent. This issue will be the subject of study in a future work. \\

 \section*{Acknowledgements}

\noindent The authors  acknowledge  CONICET (PIP: 112-201101-00325) \& UNMdP (EXA651/14) for financial support.

\newpage

\end{document}